\documentclass{iau378}

\usepackage{mathrsfs}
\usepackage{amsfonts}
\usepackage{amstext}
\usepackage{amsmath}
\usepackage{graphics,graphicx}
\usepackage{multirow}

\def\be{\begin{equation}}
\def\bea{\begin{eqnarray}}
\def\eea{\end{eqnarray}}
\def\ee{\end{equation}}
\def\tj{\theta_{\rm j}}
\newcommand\aj{\ref@jnl{AJ}}
\newcommand\araa{\ref@jnl{ARA\&A}}
\newcommand\apj{\ref@jnl{ApJ}}
\newcommand\apjl{\ref@jnl{ApJL}}     
\newcommand\apjs{\ref@jnl{ApJS}}
\newcommand\ao{\ref@jnl{ApOpt}}
\newcommand\apss{\ref@jnl{Ap\&SS}}
\newcommand\aap{\ref@jnl{A\&A}}
\newcommand\aapr{\ref@jnl{A\&A~Rv}}
\newcommand\aaps{\ref@jnl{A\&AS}}
\newcommand\azh{\ref@jnl{AZh}}
\newcommand\baas{\ref@jnl{BAAS}}
\newcommand\icarus{\ref@jnl{Icarus}}
\newcommand\jaavso{\ref@jnl{JAAVSO}}  
\newcommand\jrasc{\ref@jnl{JRASC}}
\newcommand\memras{\ref@jnl{MmRAS}}
\newcommand\mnras{\ref@jnl{MNRAS}}
\newcommand\pra{\ref@jnl{PhRvA}}
\newcommand\prb{\ref@jnl{PhRvB}}
\newcommand\prc{\ref@jnl{PhRvC}}
\newcommand\prd{\ref@jnl{PhRvD}}
\newcommand\pre{\ref@jnl{PhRvE}}
\newcommand\prl{\ref@jnl{PhRvL}}
\newcommand\pasp{\ref@jnl{PASP}}
\newcommand\pasj{\ref@jnl{PASJ}}
\newcommand\qjras{\ref@jnl{QJRAS}}
\newcommand\skytel{\ref@jnl{S\&T}}
\newcommand\solphys{\ref@jnl{SoPh}}
\newcommand\sovast{\ref@jnl{Soviet~Ast.}}
\newcommand\ssr{\ref@jnl{SSRv}}
\newcommand\zap{\ref@jnl{ZA}}
\newcommand\nat{\ref@jnl{Nature}}
\newcommand\iaucirc{\ref@jnl{IAUC}}
\newcommand\aplett{\ref@jnl{Astrophys.~Lett.}}
\newcommand\apspr{\ref@jnl{Astrophys.~Space~Phys.~Res.}}
\newcommand\bain{\ref@jnl{BAN}}
\newcommand\fcp{\ref@jnl{FCPh}}
\newcommand\gca{\ref@jnl{GeoCoA}}
\newcommand\grl{\ref@jnl{Geophys.~Res.~Lett.}}
\newcommand\jcp{\ref@jnl{JChPh}}
\newcommand\jgr{\ref@jnl{J.~Geophys.~Res.}}
\newcommand\jqsrt{\ref@jnl{JQSRT}}
\newcommand\memsai{\ref@jnl{MmSAI}}
\newcommand\nphysa{\ref@jnl{NuPhA}}
\newcommand\physrep{\ref@jnl{PhR}}
\newcommand\physscr{\ref@jnl{PhyS}}
\newcommand\planss{\ref@jnl{Planet.~Space~Sci.}}
\newcommand\procspie{\ref@jnl{Proc.~SPIE}}

\newcommand\actaa{\ref@jnl{AcA}}
\newcommand\caa{\ref@jnl{ChA\&A}}
\newcommand\cjaa{\ref@jnl{ChJA\&A}}
\newcommand\jcap{\ref@jnl{JCAP}}
\newcommand\na{\ref@jnl{NewA}}
\newcommand\nar{\ref@jnl{NewAR}}
\newcommand\pasa{\ref@jnl{PASA}}
\newcommand\rmxaa{\ref@jnl{RMxAA}}

\newcommand\maps{\ref@jnl{M\&PS}}
\newcommand\aas{\ref@jnl{AAS Meeting Abstracts}}
\newcommand\dps{\ref@jnl{AAS/DPS Meeting Abstracts}}
\begin{document}

\lefttitle{Mukesh Kumar Vyas}
\righttitle{Proceedings of the International Astronomical Union: \LaTeX\ Guidelines for~authors}

\jnlPage{0}{0}
\jnlDoiYr{2023}
\doival{0}

\aopheadtitle{Proceedings IAU Symposium}
\editors{-}

\title{Photon scattering in a relativistic outflow having velocity shear: a novel mechanism of generation for high energy power-law spectra}

\author{Mukesh Kumar Vyas, Asaf Pe'er}
\affiliation{Bar Ilan University, Ramat Gan, Israel. 5290002\\ mukeshkvys@gmail.com}

\begin{abstract}
We show that an extragalactic jet with a velocity shear gives rise to Fermi like acceleration process for photons scattering withing the shear layers of the jet. Such photons then gain energy to produce a high energy power law. These power law spectra at high energies are frequently observed in several extragalactic objects such as Gamma Ray Bursts (GRBs). We implement the model on GRBs to show that the obtained range of the photon indices are well within their observed values. The analytic results are confirmed with numerical simulations following Monte Carlo approach.
\end{abstract}


\maketitle

\section{Introduction}
Many astrophysical objects show a power law component in their high energy spectrum like gamma-ray bursts [GRBs] \citep{Band.etal.1993,Kaneko.etal.2006ApJS..166..298K,Bosmjak.etal.2014A&A...561A..25B,peer2015AdAst2015E..22P,Preece.etal.2000ApJS..126...19P,Barraud.etal.2003A&A...400.1021B}, Active galactic nuclei [AGNs] \citep{Nandra&Pounds1994,Reeves&Turner2000,Page.etal.2005} etc. They harbour a relativistic jet propagating with a Lorentz factor $\Gamma$. The existence of high energy power law in them is associated with processes like Synchrotron \cite{Band.etal.1993,Kaneko.etal.2006ApJS..166..298K} or Compton scattering \citep{1970RvMP...42..237B,1986rpa..book.....R,1993ApJ...409L..33Z, Vyas.etal.2021ApJ...908....9V, vyas.etal.2021.predictingAPJL}. 

Here we present an alternate aspect of generation of power law spectra in a relativistic flow. Considering that a relativistic jet with a differential Lorentz factor within its layers is a generic picture, we show that a photon scattering between these layers can gain or lose energy between two scattering events with electrons. However, when the photon goes through several scatterings, on average, it leads to an energy gain. A population of photons scattering within and escaping to an observer is shown to distribute like a power law spectrum at high energies. 
We argue that this mechanism is a viable candidate to generate high energy power-law spectra. Further, from the observations, this model hands us a tool to infer the jet structures and respective Lorentz factor profile. 

\section{Generation of power law spectra in relativistic jets}
Consider a population of photons injected deep inside a jet and left to scatter between shear layers before escaping. The jet assumes an angle dependent Lorentz factor profile due to shear that developed as the jet crosses the progenitor star. It is given as, 
\be 
\gamma(\theta)= \gamma_{\rm min}+\frac{\gamma_{0}}{\sqrt{\left(\frac{\theta}{\theta_{\rm j}}\right)^{2p}+1}}.
\label{eq_gamma_1}
\ee 
Here $\gamma_0$ is the maximum Lorentz factor of the jet along its axis and it decays like $\gamma \propto \theta^{-p}$ beyond $\theta > \tj$. At large angles, $\theta>>\tj$, Lorentz factor settles at $\gamma_{\rm min}$. To estimate the spectrum, we follow the following procedure to obtain energy evolution of the photons in the jet. 
After $k$ scattering $N$ photons are left within the region out of $N_0$. On average, a photon has energy $\varepsilon_k = \varepsilon_i \Bar{g}^k$ if it was injected with energy $\varepsilon_i$. Here $\Bar{g}$ is defined as the average energy gain per scattering. This implies, 
\be 
\frac{N}{N_0} = \left[\frac{\varepsilon_k  }{\varepsilon_i}\right]^\beta
\ee 
with $\beta'= \frac{\ln \bar{P}}{\ln \bar{g}}$ and the photon index $\beta$ is obtained as 
\be 
\beta=\beta'-1 = \frac{\ln \bar{P}}{\ln \bar{g}}-1.
\label{eq_photon_ind}
\ee

\begin{figure}
 \includegraphics[scale=.1] {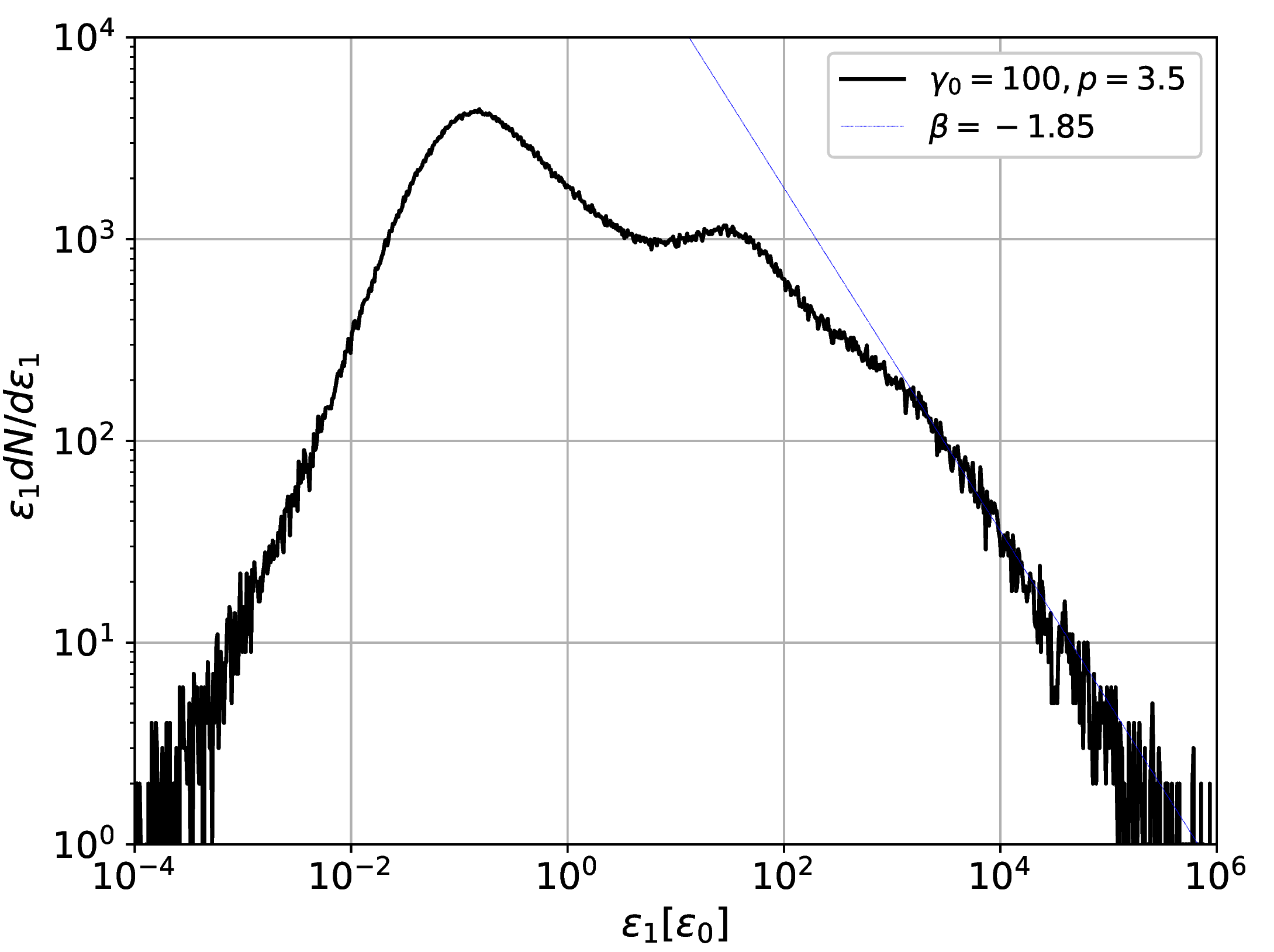}
\caption{Numerical results of the spectrum produced with parameters $\gamma_0 = 100, p = 3.5, \theta_j = 0.01$. }
\label{label_spectrum}
\end{figure}

Here $\bar{P}$ is the average probability of scattering without escape. 
The details of obtaining scattering probabilities, as well as the average energy gain for a given Lorentz factor profile are described in \cite{2023ApJ...943L...3V}. 

Alongside the analytic approach described above, we also perform numerical simulations to validate the theory. We inject photons in our simulation code that tracks and counts photons that are emitted towards certain observer with escape given escape energy. The photon population thus produce the spectrum for the escaped photons. For the details of Monte Carlo simulation code, see \cite{Vyas.etal.2021ApJ...908....9V, vyas.etal.2021.predictingAPJL, Pe'er.2008ApJ...682..463P,Lundman.etal.2013MNRAS.428.2430L}. 

\section{Results}
The photons are injected in a jet with parameters $\gamma_0 = 100, \tj = 0.01$ rad, $\gamma_{\rm min} = 1.001$ and profile index $p = 3.5$ and the produced spectrum using Monte Carlo simulations is shown in Figure 1. This spectrum is seen by an on axis observer situated at a polar angle $\theta_{\rm o} = 0$. The photons are injected with seed energy $\varepsilon_0 = 10^{-10}$ erg and scattered energies $\varepsilon_1$ are distributed in the spectra indicating a high energy power law beyond energies $\varepsilon_1/\varepsilon_0>10^{3}$. The simulated photon index $\beta = -1.85$ is confirmed with the analytic model presented in the previous section. In Figure 2, we keep all parameters to be the same and change the profile index $p$ from $2.5$ to $10$ and plot the photon indices $\beta$ with $p$. The analytic curve is shown by dashed lines while the respective simulated results are overplotted with red dots. 
The spectra gets harder as $p$ increases and saturates at $\beta = -1.55$ as $p\rightarrow\infty$. At small values of $p$, the power law spectra vanishes (for $p<2.5$). It happens due to adiabatic energy losses because the shear in jet is not sufficient to counter radial expansion and the photons do not gain energy.
The agreement between the simulations and theoretical results confirm the theoretical model developed.

\begin{figure}
\includegraphics[scale=.1] {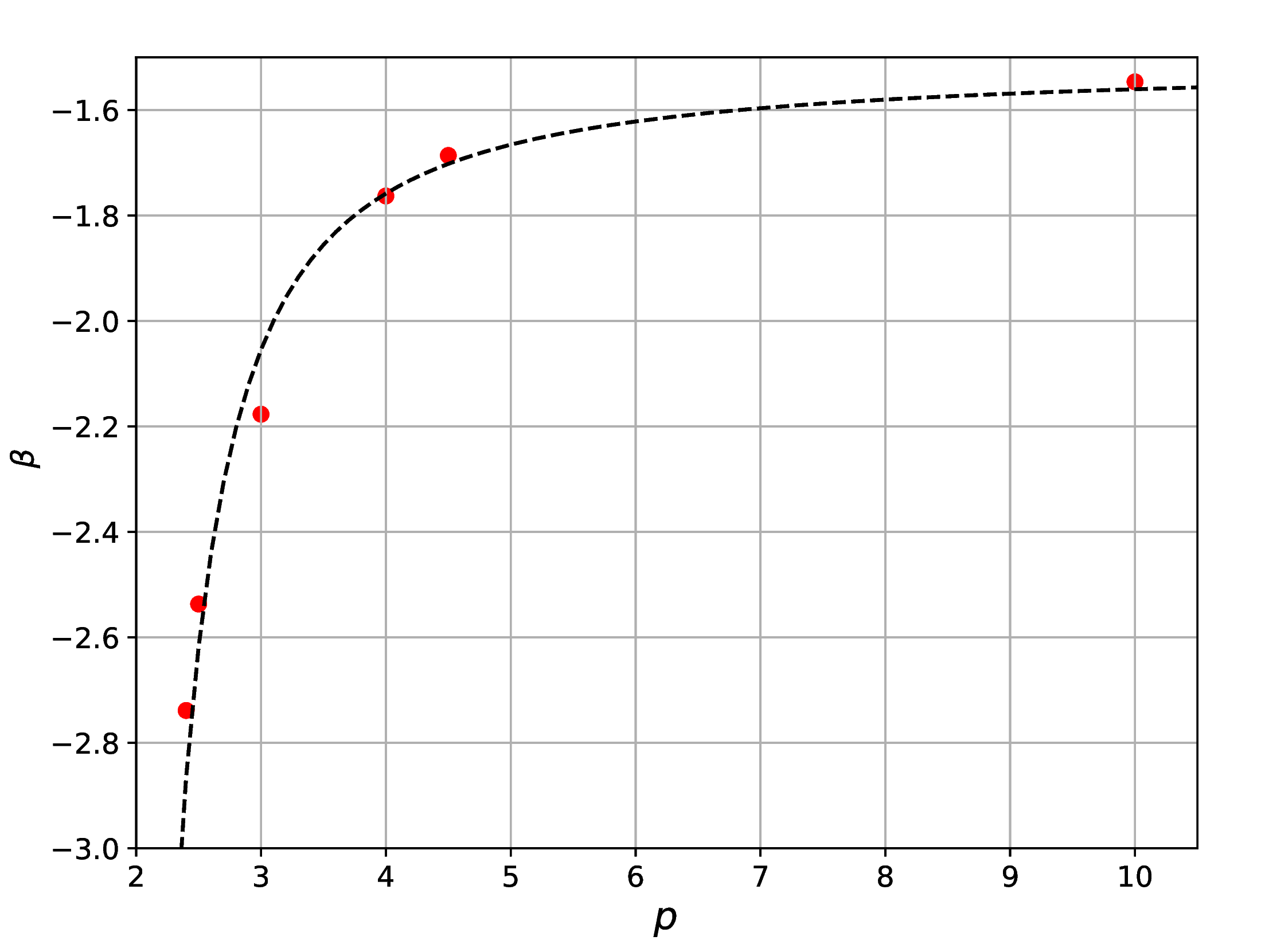}
\caption{Distribution of photon index $\beta$ with $p$ in range $2-10$. Other parameters are same as Figure 1}
\label{label_image}
\end{figure}

\section{Conclusions}
We have simulated photons that undergo scattering within the shear layers of a relativistic jet and shown that the net energy gain of these photons is capable to produce a high energy power-law spectrum. This mechanism is similar to Fermi acceleration for charged particles in the presence of magnetic fields \citep{Blandford&Eichler1987PhR...154....1B}. We discuss the same mechanism for the case of photons in shearing media with relativistic electrons. The Fermi acceleration for electrons is widely explored while we work out the photon acceleration and show that the mechanism has wide applications. The obtained range of the photon indices here is well within the range obtained in GRB observations at high energies between $-5$ to $-1.5$ \citep{Preece.etal.1998ApJ...496..849P, Kaneko.etal.2006ApJS..166..298K}.

In further works, we will work out the implications of this mechanism in other astrophysical objects such as jets in active galactic nuclei (AGNs) and accretion discs around black holes.


\end{document}